\documentclass[aps,superscriptaddress,twocolumn,prd]{revtex4-2}

\usepackage{bm,graphicx,textcomp,amssymb,amsmath,dcolumn,color}

\newcommand{\ket}[1]{\left|#1\right>}
\newcommand{\bra}[1]{\left<#1\right|}

\newcommand{\nn}{\nonumber\\}

\newcommand{\f}[1]{\mbox{\boldmath$#1$}}

\newcommand{\bea}{\begin{eqnarray}} 
\newcommand{\ea}{\end{eqnarray}}
\newcommand{\eea}{\end{eqnarray}}
\newcommand{\ord}{\,{\cal O}}

\begin{document}

\title{Magneto-photoelectric effect in graphene via tailored 
potential landscapes}

\author{Joris Josiek}

\affiliation{Helmholtz-Zentrum Dresden-Rossendorf, 
Bautzner Landstra{\ss}e 400, 01328 Dresden, Germany,}

\affiliation{Institut f\"ur Theoretische Physik, 
Technische Universit\"at Dresden, 01062 Dresden, Germany,}

\author{Friedemann Queisser}

\affiliation{Helmholtz-Zentrum Dresden-Rossendorf, 
Bautzner Landstra{\ss}e 400, 01328 Dresden, Germany,}

\affiliation{Institut f\"ur Theoretische Physik, 
Technische Universit\"at Dresden, 01062 Dresden, Germany,}

\author{Stephan Winnerl} 

\affiliation{Helmholtz-Zentrum Dresden-Rossendorf, 
Bautzner Landstra{\ss}e 400, 01328 Dresden, Germany,}

\author{Ralf Sch\"utzhold}

\affiliation{Helmholtz-Zentrum Dresden-Rossendorf, 
Bautzner Landstra{\ss}e 400, 01328 Dresden, Germany,}

\affiliation{Institut f\"ur Theoretische Physik, 
Technische Universit\"at Dresden, 01062 Dresden, Germany,}

\date{\today}

\begin{abstract}
We consider the propagation of charge carriers in planar graphene under the 
combined influence of a constant transversal magnetic field $B$ and an 
in-plane varying electric potential $\phi(x)$. 
By suitably designing the potential landscape $\phi(x)$, we may effectively 
steer charge carriers generated by photo-excitation, for example, in order 
to achieve an efficient charge separation.  
These finding may pave the way for transport schemes or 
photoelectric/photovoltaic applications. 
\end{abstract}

\maketitle

\section{Introduction}  

The special properties of graphene, such as the high charge carrier mobility
and their long mean free path, suggest utilizing them for photoelectric or 
photovoltaic applications 
\cite{WS20,MW21,PC13,MN08,SH10,BR09,LC14,ZL13,BS10,BT13,KB23,
WB20,GR20,SS18,TP15,DP19,RI19,CH15,MW18,WC09}.
In contrast to semiconductors, for example, graphene has no intrinsic band gap.
On the one hand, this can be advantageous since restrictions such as the 
Shockley-Queisser limit do not necessarily apply \cite{SQ61}. 
On the other hand, it requires alternative schemes for charge separation,
see also \cite{QS13,SK17}.
Obviously, one has to break the 
$\mathfrak C$ (charge),
$\mathfrak P$ (parity) and 
$\mathfrak T$ (time reversal) 
symmetries in order to achieve charge separation.
One way to do so is to apply an external magnetic field, but this is not 
enough.
In order to turn the usual circular cyclotron orbits in a magnetic field 
into a directed motion of the charge carriers, one can use geometric 
constraints such as folded graphene  \cite{QS13} or a graphene edge 
\cite{SK17,QL23,JC11}, for example. 
However, such geometric constraints go along with experimental and 
technological challenges. 
In order to avoid these difficulties, we consider an alternative scheme 
based on an additional electrostatic potential $\phi(\f{r})$. 
This could be induced externally by electrodes exhibiting gate voltages 
internally or by effectively doping the graphene sheet, which shifts the 
local chemical potentials, see also
\cite{OH08,SA11,SA07,WN08,KS20,BB22,PL10,WD07}.
In this way, one can systematically manipulate the non-equilibrium 
dynamics of the charge carriers generated via photo-excitation, 
for example. 

\section{Dirac Equation}  

On length scales far above the lattice spacing of approximately $0.25~\rm nm$ 
and energies well below the hopping energy $\approx2.8~\rm eV$, we may describe 
electrons and holes by an effective Dirac equation in 2+1 dimensions $(\hbar=1)$ 
\begin{eqnarray}
\label{Dirac}
\gamma^\mu\left(\partial_\mu+iqA_\mu\right)\Psi=0 
\,,
\end{eqnarray}
with $x^\mu=[v_{\rm F}t,x,y]$, where $v_{\rm F}\approx10^6\rm m/s$ 
is the Fermi velocity \cite{CG09}.
The Dirac matrices $\gamma^\mu=[\sigma^z,i\sigma^y,-i\sigma^x]$ 
acting on the spinor $\Psi=[\psi_1,\psi_2]$
are related to the Pauli matrices $\sigma^{x,y,z}$.
In this representation, the Dirac $\alpha$ and $\beta$ matrices are given 
by $\alpha^x=\sigma^x$, $\alpha^y=\sigma^y$ and $\beta=\sigma^z$. 
The vector potential $A_\mu=[\phi(x),0,Bx]$ in the Landau gauge  
generates the electric field $E(x)$ in $x$-direction and the 
magnetic field $B$ in $z$-direction.  

In view of the translation symmetry in $t$ and $y$, 
we can make the usual separation ansatz for the modes 
\bea
\Psi(t,x,y)=\exp\left\{-i\omega t+iky\right\}\,\Psi^{\omega,k}(x)
\,,
\ea
arriving at the two coupled equations 
\begin{eqnarray}
\label{Dirac-coupled}
v_\mathrm{F}[\partial_x+k+qBx]\psi^{\omega,k}_2(x)
&=& 
i[\omega-q\phi(x)]\psi^{\omega,k}_1(x)
\,,
\nn
v_\mathrm{F}[\partial_x-k-qBx]\psi^{\omega,k}_1(x)
&=& 
i[\omega-q\phi(x)]\psi^{\omega,k}_2(x)
\,.
\end{eqnarray}
Without loss of generality, we choose $\psi^{\omega,k}_1(x)$ 
to be real while $\psi^{\omega,k}_2(x)$ is imaginary.  

\subsection{Dispersion Relation}   

\begin{figure}[ht]
\includegraphics[width=.9\columnwidth]{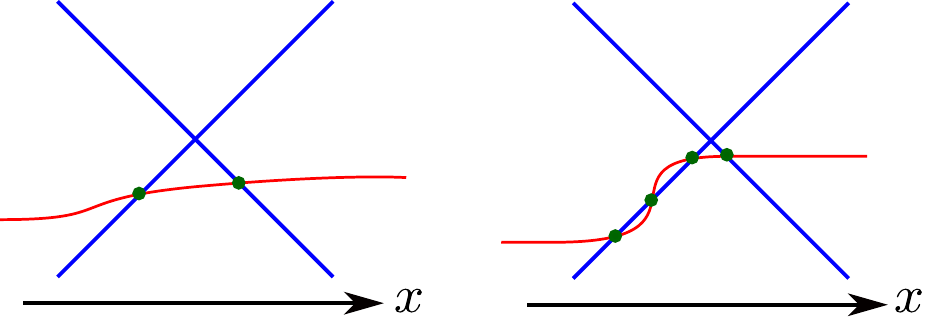} 
\caption{Sketch of solutions to the dispersion relation for the 
sub-threshold case (left) and the super-threshold case (right). 
The red 
curve denotes the potential $q\phi(x)$ while the black 
lines correspond to $\omega\pm v_\mathrm{F}[k+qBx]$.
In the sub-threshold case (left), the classically allowed solution 
lies in between the two intersections with the two straight lines, 
which is referred to as type I. 
In the super-threshold case (right), there is an additional 
classically allowed solution lying in between two intersections 
with the {\em same} straight line, which is referred to as type II.}
\label{figure-disp}
\end{figure}

In order to obtain a first insight into the structure of the solutions to 
Eq.~\eqref{Dirac-coupled}, we employ WKB analysis and arrive at the 
dispersion relation 
\bea
\label{dispersion}
p_x^2(x)+[k+qBx]^2=\frac{[\omega-q\phi(x)]^2}{v_\mathrm{F}^2}
\,,
\ea
where $p_x(x)$ denotes the momentum in $x$-direction. 
The classical turning points where $p_x=0$ can thus be obtained from the 
intersections between the potential curve $\phi(x)$ and straight lines 
with slope $\pm v_\mathrm{F}B$, shifted by $k$ and $\omega$, 
i.e., $q\phi(x)=\omega\pm v_\mathrm{F}[k+qBx]$, see 
Fig.~\ref{figure-disp}. 

\subsection{Sub-threshold Fields}\label{Sub-threshold Fields}   

Assuming that the potential $\phi(x)$ approaches constant values 
asymptotically $\phi(x\to\pm\infty)=\rm const$, the possible number 
of intersections depends on the maximum slope of $\phi(x)$, 
i.e., the electric field $E(x)$. 
If this field is sub-threshold $E(x)<v_\mathrm{F}B$ everywhere, we always 
find two intersections -- one for each straight line, 
see Fig.~\ref{figure-disp} left. 
In the following, we refer to these solutions as type I. 

\begin{figure}[ht]
\includegraphics[width=.9\columnwidth]{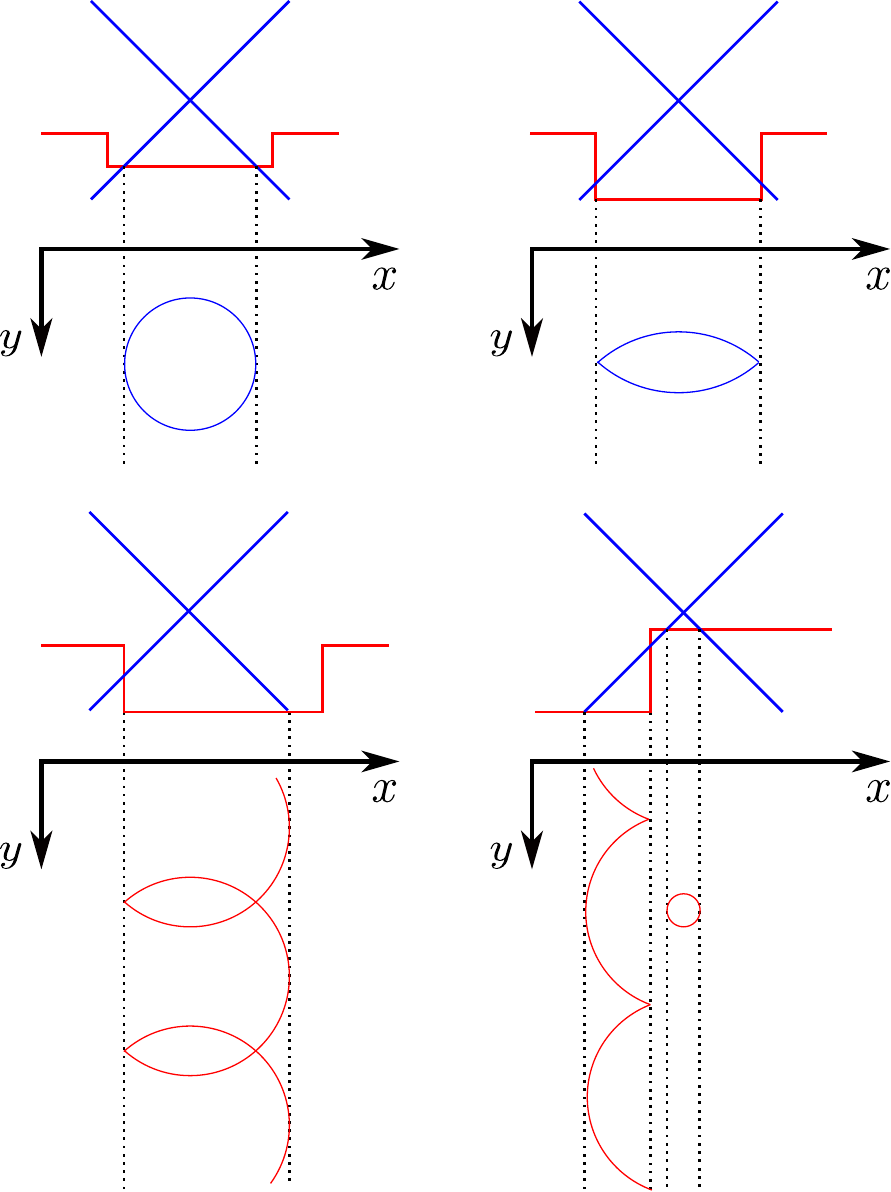}
\caption{Examples of semi-classical trajectories for various solutions 
of the dispersion relation. 
The upper left panel illustrates cyclotron motion, driven exclusively 
by the magnetic field. 
In contrast, the upper right panel displays a localized trajectory, 
with turning points defined by reflections at the potential $\phi(x)$.
A mixed case is depicted in the lower left panel. 
So far, all these solutions are of type I, causing the propagation 
velocity in the $v_y$-direction to change sign along the trajectory. 
Finally, the lower right panel demonstrates the coexistence of a 
type II solution (where $v_y\neq0$) and a localized cyclotron motion.} 
%
%
\label{types}
\end{figure}

In the limit of a vanishing electric field $\phi=\rm const$, 
this case just corresponds to the usual circular (cyclotron) motion, 
see Fig.~\ref{types} top left. 
The distance between the two intersections is then twice the cyclotron 
radius 
\bea
\label{cyclotron}
r_{\rm cyc}=\frac{\omega-q\phi}{qBv_\mathrm{F}}
\,.
\ea
Adding a small and constant electric field induces a drift velocity 
in orthogonal direction $\f{v}_{\rm drift}=\f{E}\times\f{B}/B^2$. 
Exploiting the quasi-relativistic form of Eq.~\eqref{Dirac}, this can be 
understood via applying an effective Lorentz boost with $\f{v}_{\rm drift}$ 
after which the electric field in the co-moving frame vanishes. 
However, since this drift velocity is the same for particles and holes, 
it cannot be used to achieve charge separation directly. 

\subsection{Super-threshold Fields}\label{Super-threshold Fields}   

Let us now consider an electric field which exceeds the threshold 
$E(x)>v_\mathrm{F}B$ is some region. 
In the limiting case of constant fields with $E>v_\mathrm{F}B$, an 
effective Lorentz boost could transform the magnetic field away. 

Note that the same condition $E>v_\mathrm{F}B$
is required for observing the graphene 
analogue of the Sauter-Schwinger effect, i.e., electron-positron pair 
creation of the vacuum in quantum electrodynamics (QED) induced by a 
strong (and slowly varying) electric field \cite{Sau31,Sch51}.
However, this field strength $E>v_\mathrm{F}B$ should not be confused 
with the analogue of the Schwinger critical field in QED 
\bea
\label{Schwinger}
E_{\rm crit}=\frac{m_e^2c^3}{\hbar q}\approx1.3\times10^{18}\,
\frac{\rm V}{\rm m}
\,,
\ea
where $m_e$ is the electron mass and $c$ the speed of light. 
Since graphene does not have a band gap, there is no direct analogue to 
this quantity~\eqref{Schwinger}.
However, as one may infer from Eq.~\eqref{dispersion}, for example, 
a transversal momentum $k_\perp$ could act like an effective gap 
$v_\mathrm{F}k_\perp$. 
If we assume such a gap in the eV regime, the graphene analogue of the 
above critical field~\eqref{Schwinger} would be of order $10^9~\rm V/m$.
In contrast, even for a magnetic field of one Tesla, the field strength 
required for reaching $E>v_\mathrm{F}B$ is much lower 
$E=\ord(10^6~\rm V/m)$.
Generating such a field -- e.g., via a  potential difference 
$\Delta\phi=\ord(\rm V)$ over a sub-micrometer scale -- is much 
easier than reaching $10^9~\rm V/m$.

For super-threshold fields $E(x)>v_\mathrm{F}B$, there may be additional 
intersections, which now lie on the {\em same} straight line,
as in Fig.~\ref{figure-disp} right. 
These additional solutions will be referred to as type II. 

Since both intersections lie on the same straight line
(e.g., the line with positive slope in Fig.~\ref{figure-disp} right), 
the propagation velocity in $y$-direction, 
which is determined by the mechanical momentum $k+qBx$, does not 
change sign in between these turning points -- which means that 
the motion in $y$-direction is not reversed (in contrast to the 
circular orbits discussed above). 
As an intuitive picture, one turning point is caused by the bending due 
to the magnetic field, while the other turning point is caused by a 
reflection at the electrostatic potential barrier. 

As one can already infer from Fig.~\ref{figure-disp}, one can have a 
type-I solution without a type-II solution, but not the other way around
(though they may lie on different branches). 
In general, one can have a coexistence of the different trajectories, 
as in Fig.~\ref{types} bottom right. 



\subsection{Quantum Solutions}   

On the classical level, the case of two intersections with $p_x^2>0$ 
in between corresponds to a bound state in $x$-direction which could,
however, display a non-zero average velocity in $y$-direction. 
If the size $L$ of the classically allowed region where $p_x^2>0$
is large enough, i.e., $p_x^2L^2\geq\ord(\hbar^2)$, these bound 
states will also exist on the quantum level, i.e., as solutions to 
Eq.~\eqref{Dirac-coupled}.

The associated stationary states will be localized in $x$-direction
(with exponential tales in the classically forbidden region $p_x^2<0$)
and plane waves in $y$-direction, characterized by a discrete set of 
frequencies $\omega(k)$.
The motion in $y$-direction can either be obtained from the group 
velocity $d\omega/dk$ or the current -- where we can apply many of the 
arguments already discussed in \cite{QS13} after incorporating the 
additional potential $\phi(x)$. 

To be more explicit, let us consider the current $j_{\omega,k}^\mu$ 
associated to a eigen-solution $\Psi^{\omega,k}$ which reads 
\bea
j_{\omega,k}^\mu=v_\mathrm{F}\bar\Psi^{\omega,k}\gamma^\mu\Psi^{\omega,k}
\,.
\ea
The density $\rho=|\psi^{\omega,k}_1|^2+|\psi^{\omega,k}_2|^2$ is recovered 
for $\mu=0$. 
As expected for stationary solutions which are 
localized in $x$-direction, the $\mu=1$ contribution, 
i.e., the current 
$j_{\omega,k}^x=v_\mathrm{F}\Psi^{\omega,k}\sigma^x\Psi^{\omega,k}=
v_\mathrm{F}(\psi^{\omega,k}_1)^*\psi^{\omega,k}_2+\rm h.c.=0$ 
vanishes identically since $\psi^{\omega,k}_1$ is real while 
$\psi^{\omega,k}_2$ is imaginary. 

The remaining current in $y$-direction 
\bea
j_{\omega,k}^y=v_\mathrm{F}\Psi^{\omega,k}\sigma^y\Psi^{\omega,k}
=
iv_\mathrm{F}(\psi^{\omega,k}_2)^*\psi^{\omega,k}_1+\rm h.c. 
\ea
can be simplified to 
$j_{\omega,k}^y=-2iv_\mathrm{F}\psi^{\omega,k}_1\psi^{\omega,k}_2$.
Assuming that $\omega$ lies outside the range of $q\phi(x)$, 
as in Fig.~\ref{figure-disp}, we may derive the total current 
in $y$-direction by inserting Eq.~\eqref{Dirac-coupled} and 
integrating over $x$, which gives the two equivalent expressions 
\bea
J_{\omega,k}^y
&=&
\int dx\,j_{\omega,k}^y
\nn
&=& 
2v_\mathrm{F}^2\int dx
\left[
\frac{k+qBx}{\omega-q\phi}+
\frac{q\phi'}{2[\omega-q\phi]^2}
\right] 
\left|\psi^{\omega,k}_1\right|^2
\nn
&=& 
2v_\mathrm{F}^2\int dx
\left[
\frac{k+qBx}{\omega-q\phi}-
\frac{q\phi'}{2[\omega-q\phi]^2}
\right] 
\left|\psi^{\omega,k}_2\right|^2
.\quad
\ea
The second term $\propto q\phi'$ stems from the commutator of 
$\partial_x$ and $q\phi(x)$ and can be neglected in the 
semi-classical limit $p_x^2L^2\gg\hbar^2$. 
Furthermore, it comes with opposite signs depending on the 
representation (in terms of $\psi^{\omega,k}_1$ or $\psi^{\omega,k}_2$).
Thus, the direction (i.e., sign) of the current $J_{\omega,k}^y$ is 
determined by the first term, which is the same in both representations. 

\subsection{Charge Separation}   

To infer the direction of the current $J_{\omega,k}^y$, we have to 
study the relative sign between $k+qBx$ and $\omega-q\phi(x)$. 
For type-I solutions, the sign of $k+qBx$ changes in between the turning 
points such that the integral could be positive or negative or even 
zero (which is the case for circular orbits in a pure magnetic field).
For type-II solutions, on the other hand, $k+qBx$ does not change sign 
in the classically allowed region. 

As a result, the associated bound states in $x$-direction move with a 
finite velocity in $y$-direction (assuming that one can neglect the 
exponentially suppressed contributions of the evanescent tails).
Hence, these modes are very relevant for achieving charge separation.
Even if $\omega$ lies inside the range of $q\phi(x)$, we may draw the 
same conclusion on the semi-classical level, because the point where 
$\omega=q\phi(x)$ is also outside the classically allowed region. 

The above conclusion based on the current can also be applied to the 
group velocity $d\omega/dk$.
To show this, let us write Eq.~\eqref{Dirac-coupled} in terms of the 
Dirac Hamiltonian $\hat H_k$ per mode $k$ in the form 
$\hat H_k\ket{\Psi^{\omega,k}}=\omega\ket{\Psi^{\omega,k}}$ where 
\bea
\hat H_k
=
\left(
\begin{array}{cc}
q\phi(x) & -iv_\mathrm{F}[\partial_x+k+qBx]
\\
-iv_\mathrm{F}[\partial_x-k-qBx] & q\phi(x)
\end{array}
\right) 
.
\qquad
\ea
Taking the $k$-derivative $d\hat H_k/dk=v_\mathrm{F}\sigma^y$, we see that 
the current $J^y$ 
can be written as 
$\bra{\Psi^{\omega,k}} d\hat H_k/dk \ket{\Psi^{\omega,k}}$. 
Next, taking the $k$-derivative of 
$\hat H_k\ket{\Psi^{\omega,k}}=\omega\ket{\Psi^{\omega,k}}$,
we find that the current $J^y$ coincides with the group velocity 
$d\omega/dk$ for normalized solutions $\ket{\Psi^{\omega,k}}$. 

The triangle inequality $2|ab|\leq|a|^2+|b|^2$ implies that 
$|j^y|\leq v_\mathrm{F}\rho$, i.e., the group velocity $d\omega/dk$
cannot exceed the Fermi velocity $v_\mathrm{F}$, as expected. 

\section{Semi-classical Trajectories}  

\begin{figure}[ht]
\includegraphics[height=0.44\textheight]{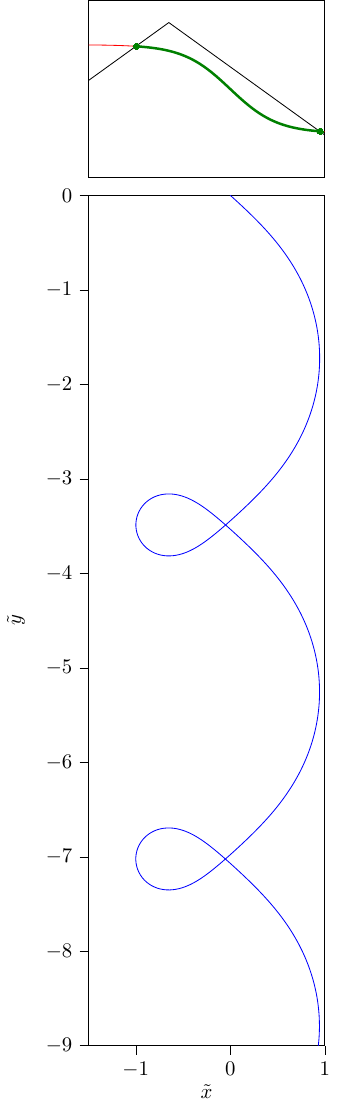} 
\includegraphics[height=0.44\textheight]{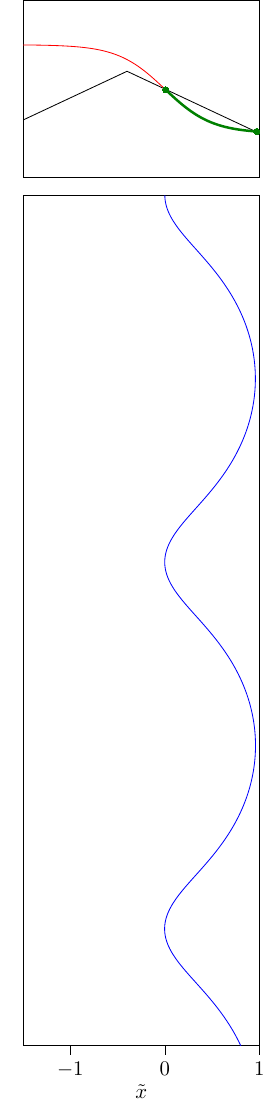} 
\includegraphics[height=0.44\textheight]{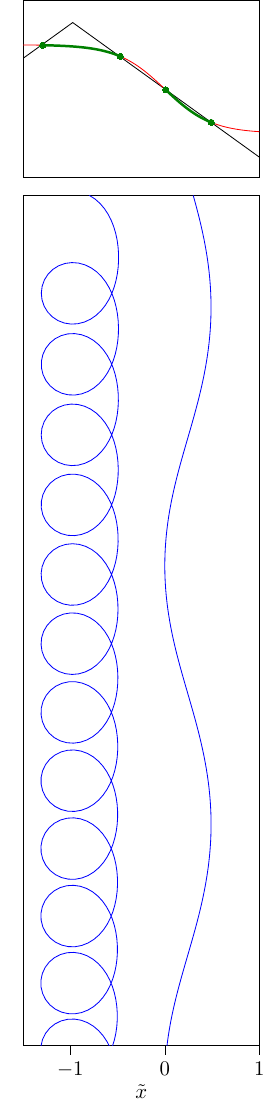} 
\caption{Examples for solutions of the dispersion relation (top) 
and the associated trajectories (bottom). 
The left panel shows a type-I solution where the $y$-velocity $v_y$
is opposite at the two turning points where $v_x=0$ and thus vanishes 
somewhere in between. 
For a type-II solution as shown in the middle panel, on the other hand, 
the motion in $y$-direction is never reversed, only $x(t)$ displays 
turning points. 
The right panel depicts the coexistence of a type-I and a type-II solution.}
\label{figure-tra}
\end{figure}

In order to visualize the type I and II solutions discussed above, let us 
study the associated semi-classical trajectories $\f{r}(t)=[x(t),y(t)]$.
Since the effective Dirac equation~\eqref{Dirac} does not contain a mass 
term, we are in the ultra-relativistic limit where the velocity is given by 
\bea
\dot{\f{r}}=v_\mathrm{F}\,\frac{\f{p}}{|\f{p}|}
\,,
\ea
with the momentum $\f{p}$ satisfying the equation of motion 
\bea
\dot{\f{p}}=q\left(\f{E}+\dot{\f{r}}\times\f{B}\right) 
\,.
\ea
Solving this coupled set of equations numerically yields the 
trajectories $\f{r}(t)$ where some example cases are depicted 
in Fig.~\ref{figure-tra}. 

The group velocity $d\omega/dk$ or the current $J^y$ of the stationary 
solutions $\Psi^{\omega,k}$ is then related to the average velocity 
$v_y$ on the classical level. 
For the type-II solutions, the velocity $v_y(t)$ displays some 
time-dependence but never changes its sign -- such that the average 
velocity $v_y$ is also non-zero. 
For the type-I solutions, on the other hand, the velocity $v_y(t)$
does change its sign during each cycle. 
However, we found that most of them do also yield a non-vanishing 
average velocity $v_y$. 
Only in quite special cases, it averaged out to zero. 

To understand the origin of this average velocity $v_y$, let us consider
the simplified case of rectangular potential barriers, see 
Figs.~\ref{oddpot} and \ref{evenpot}. 
We see that the average velocity $v_y$ only vanishes for the symmetric 
scenarios where the contributions from the two branches 
$\omega\pm v_\mathrm{F}[k+qBx]$ exactly cancel each other. 





\section{Symmetries}  

Now let us investigate the solutions discussed above in light of the 
$\mathfrak C$, $\mathfrak P$ and $\mathfrak T$ symmetries already 
mentioned in the Introduction.
Applied separately, each of these transformations would reverse the 
electric and/or magnetic field and thus change the set-up under 
consideration. 
Thus, we shall try to find a combined symmetry which keeps the electric 
and magnetic fields fixed. 
To this end, we have to specify how the potential $\phi(x)$ 
transforms under reflection $x\to-x$. 

\subsection{Odd Potential}  

\begin{figure}[ht]
\includegraphics[width=6cm]{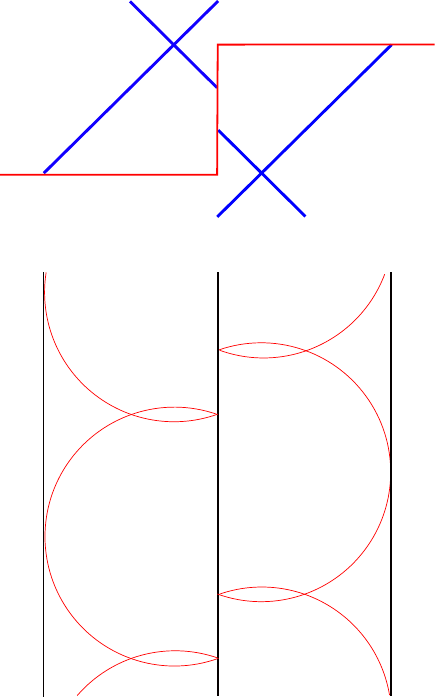}
\caption{Example of an odd potential $\phi(x)$ which permits the existence 
of paired electron and hole solutions that propagate in the same $y$ direction.}
\label{oddpot}
\end{figure}

First, we assume that the potential is an odd function $\phi(-x)=-\phi(x)$
such as a potential step, see Fig.~\ref{oddpot}.
Then we find that Eq.~\eqref{Dirac-coupled} remains invariant under the  
combined reflection $x\to-x$ as well as $k\to-k$ and $\omega\to-\omega$.
Since both $k$ and $\omega$ change sign, the group velocity
(along the potential, i.e., in $y$-direction) stays the same. 

For example, if we have an electron solution with $\omega>0$ 
(assuming a vanishing chemical potential $\mu=0$) which is localized
on the right-hand side of potential step and moves in $y$-direction, 
the above symmetry implies the existence of a hole solution $\omega<0$ 
which is localized on the left-hand side of potential step and moves 
in the same $y$-direction, see Fig.~\ref{oddpot}.

\subsection{Even Potential}  

\begin{figure}[ht]
\includegraphics[width=6cm]{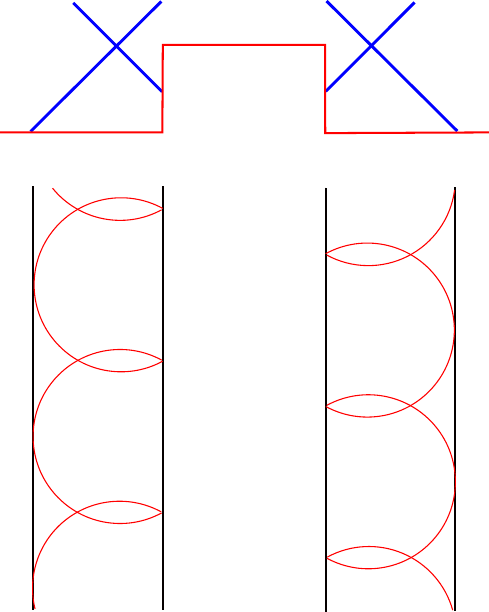}
\caption{Example of an even potential $\phi(x)$ which permits the existence 
of paired solutions with the same frequency (either both electron solutions 
or both hole solutions) that propagate in opposite directions.}
\label{evenpot}
\end{figure}

Now let us assume an even function $\phi(-x)=\phi(x)$ such as a potential
barrier, see Fig.~\ref{evenpot}.
In this case Eq.~\eqref{Dirac-coupled} remains invariant if we apply the 
reflection $x\to-x$ as well as $k\to-k$ (but now $\omega$ remains fixed) 
and simultaneously transform the spinor wave-function via 
$\psi^{\omega,k}_2\to-\psi^{\omega,k}_2$, i.e., 
$\Psi^{\omega,k}\to\sigma^z\Psi^{\omega,k}=(\Psi^{\omega,k})^*$.

In this case, a solution which is localized on the right-hand side of 
potential barrier and moves in $y$-direction, for example, 
is transformed into another solution with the same frequency 
which is localized on the left-hand side of potential barrier 
and moves into the opposite $y$-direction, see Fig.~\ref{evenpot}.

At the same time, a solution with the opposite frequency can move 
on a qualitatively different trajectory, e.g., it could experience the 
potential barrier as a potential dip which can be overcome easily. 

\section{Magneto-photoelectric Effect}  

Now we have all the ingredients required for designing a potential landscape
which facilitates an effective charge separation. 
First, we found a definite propagation direction 
(at least for all type-II solutions, but also for most of the type-I cases)
orthogonal to the electric and magnetic fields (i.e., in $y$-direction).
Second, the trajectories are different for electrons and holes.
Apart from the opposite orientation of their (quasi) circular orbits in the 
magnetic field, a negative potential barrier $\phi(x)<0$ acts as a reflector 
for electrons (as long as it is high and wide enough) while holes can 
traverse it -- while it is the other way around for a positive potential 
barrier $\phi(x)>0$. 

\begin{figure}
\includegraphics[width=\columnwidth]{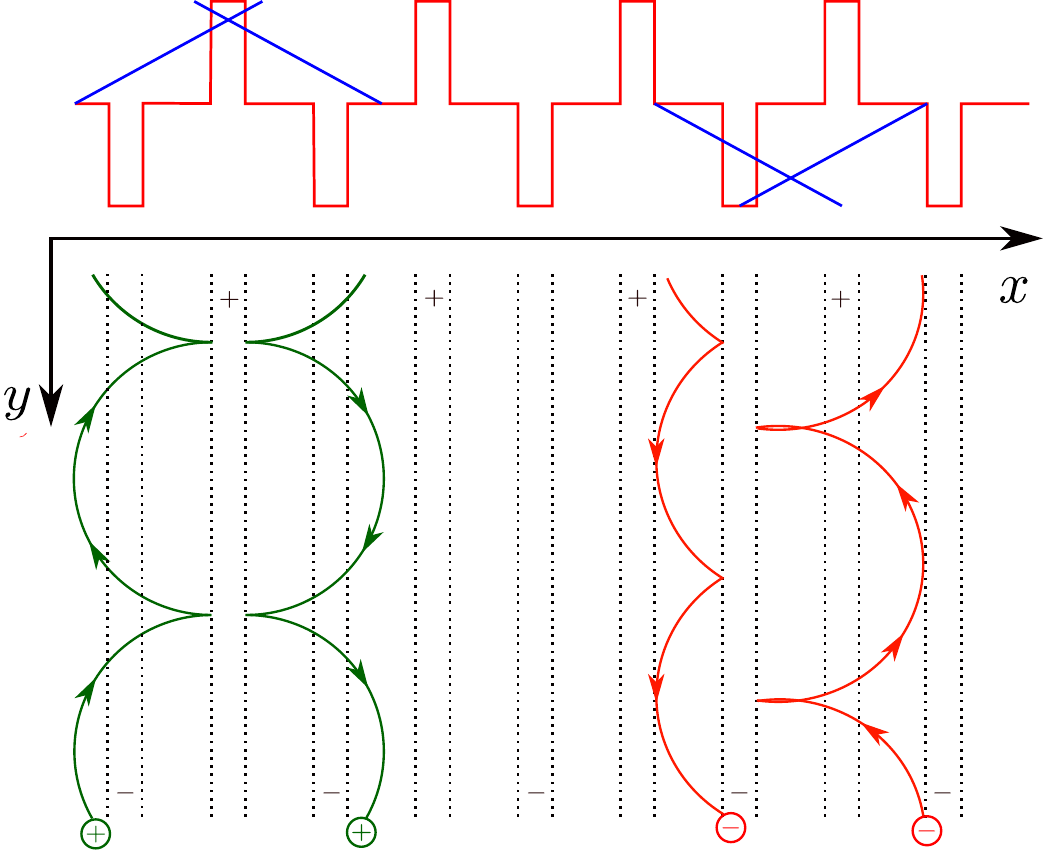}
\caption{Arranging positive and negative potential barriers in a double comb
like structure allows us to separate the charge carriers. 
Negative potential barriers reflect electrons but can be traversed by holes.
Conversely, holes are reflected by positive potential barriers while being 
able to propagate through negative potentials.} 
\label{comb}
\end{figure}

An example for such a potential landscape $\phi(x)$ is sketched in Fig.~\ref{comb}.
By comb-like structures of alternating positive and negative potential 
barriers, we can steer the charge carriers. 
The majority of the electron trajectories (with large enough cyclotron radii) 
encircle the negative potential barriers in counter-clockwise direction 
while the hole trajectories encircle the positive potential barriers 
in clockwise direction.

\begin{figure}[ht]
\includegraphics[width=.7\columnwidth]{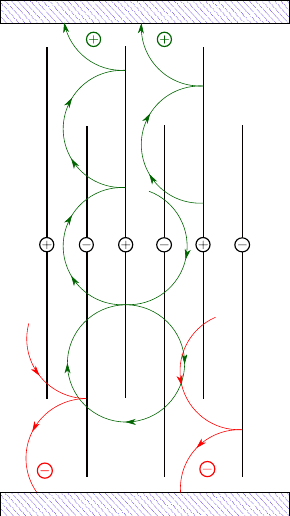}
\caption{Sketch of a geometry in order to collect the positive and 
negative charge carriers at the top and bottom metallic contacts, 
respectively.
The negative potential barriers which are reflectors for the
electrons are extended towards the negative $y$-direction while 
the positive potential barriers are extended upwards.}
\label{contacts}
\end{figure}

To collect the charge carriers, one could imagine extending the positive 
potential barriers a bit further in positive $y$-direction than the 
negative ones (i.e., breaking the translational invariance in $y$-direction 
at some point) and attaching electric contacts to the graphene which absorb 
the arriving holes.
The electrons could be collected by an analogous procedure on the other side,
i.e., in negative $y$-direction.
A sketch of such a set-up is depicted in Fig.~\ref{contacts}.
Alternatively, one could collect the electrons and holes by additional 
(negative and positive) potential barriers parallel to the $x$-axis, 
which terminate the other potential barriers 
(forming two combs telescoped into each other) and steer the electrons 
and holes to the contacts. 

\section{Experimental Parameters}  

In order to discuss the order of magnitude of the involved parameters,
let us first consider charge carriers of rather high energy 
$\omega=\ord(\rm eV)$ which could be created by the absorption of 
optical photons, for example.
Then, assuming a magnetic field of one Tesla, we obtain a cyclotron 
radius~\eqref{cyclotron} of order $\mu\rm m$, i.e., comparable to the 
mean free path in graphene. 
If the mean free path would be much smaller than the cyclotron 
radius~\eqref{cyclotron}, the directed motion of the charge carriers 
would be suppressed by permanent scattering events. 
Furthermore, the cyclotron radius~\eqref{cyclotron} is much larger than 
the Landau length $\ell_{B}=\sqrt{\hbar/(qB)}\approx26~\rm nm$ which shows 
that the classical picture should provide a good approximation. 

As already explained in Sec.~\ref{Super-threshold Fields}, electric fields 
exceeding the threshold of $10^6~\rm V/m$ or $\rm V/\mu m$ can be 
generated by potential differences $\Delta\phi=\ord(\rm V)$ over 
sub-micrometer length scales. 
As one example, one could consider a graphene sheet on a boron-nitrate 
layer of thickness between 10 and 100~nm.
Via lithography, electrodes of around 200~nm width could be placed on 
the other side of the insulating boron-nitrate layer 
\cite{KM13,LK12,LG17}.
In this way, the potential barrier height and width should be large enough 
to ensure efficient reflection.  

Furthermore, as became evident in the previous section, the distance between 
the potential barriers should be of same order as the cyclotron 
radius~\eqref{cyclotron} determined by the magnetic field $B$ and the energy 
of the charge carriers. 
Thus, by decreasing the applied magnetic field, one could adapt the scheme 
to lower energies of the charge carriers 
(i.e., lower frequencies of the incident radiation). 
Alternatively, one could imagine switching on or off electrodes in order 
to match the distance between the active electrodes accordingly. 

\bigskip 

\section{Conclusion and Outlook}  

By solving the effective Dirac equation (valid in the vicinity of the 
Dirac cones), we study the motion of charge carriers in planar graphene 
under the combined influence of a constant perpendicular magnetic field $B$ 
and an in-plane electrostatic potential landscape $\phi(x)$, which 
could be generated by external electrodes or by doping, for example.  
Shaping the potential landscape $\phi(x)$ and thereby breaking the 
$\mathfrak C$ (charge),
$\mathfrak P$ (parity) and 
$\mathfrak T$ (time reversal) 
symmetries in an appropriate way, we may effectively steer the charge 
carriers and achieve charge separation.
This could be relevant for graphene based photoelectric or photovoltaic 
applications or for graphene electronics etc. 

Since the directed motion of charge carriers in the scenarios considered 
here already follows from general principles, we expect that the steering 
phenomenon is quite robust against small perturbations and imperfections, 
which should help us to realize this effect experimentally. 
One effect which we have not considered here is the (Auger like) creation 
of secondary electron-hole pairs via the strong Coulomb interaction, 
see also \cite{QL23}. 
Since such effects are expected to be most pronounced at the edge of 
graphene (or other imperfections such as defects) which are not part 
of our set-up, they might be less relevant here than in other scenarios 
(see, e.g., \cite{SK17}), but this should be the subject of further studies. 

\bigskip


\acknowledgments 

Funded by the Deutsche Forschungsgemeinschaft 
(DFG, German Research Foundation) -- Project-ID 278162697-- SFB 1242.

\end{document}